\documentclass[useAMS,usenatbib]{mn2e}
\usepackage{epsfig}
\usepackage{times}

\newcommand{\com}[1] {#1}
\newcommand{\etal}{et al.}

\newcommand{\rmd}{{\rm d}}

\title[Enlarged Cosmic Shear Survey with the WHT]
{An Enlarged Cosmic Shear Survey with the William Herschel Telescope}

\author[R. Massey \etal]
{Richard~Massey,$^{1,2}$\thanks{E-mail: {\tt rjm@astro.caltech.edu}}
Alexandre~Refregier,$^3$
David~J.~Bacon,$^4$
Richard~Ellis$^2$ and 
Michael~L.~Brown$^4$\\
$^1$ Institute of Astronomy, Madingley Road, Cambridge CB3 OHA\\
$^2$ California Institute of Technology, Pasadena, California 91125, USA\\
$^3$ Service d'Astrophysique, CEA Saclay, F-91191 Gif sur Yvette, France\\
$^4$ Institute for Astronomy, Blackford Hill, Edinburgh EH9 3HJ\\}

\date{Accepted ---. Received ---; in original form 8 April 2004.}

\pagerange{\pageref{firstpage}--\pageref{lastpage}}
\pubyear{2004}

\begin{document}

\maketitle

\label{firstpage}

\begin{abstract}  We report the results of a cosmic shear survey using the 4.2m
William Herschel Telescope on La Palma, to a depth of $R=25.8$
($z\approx0.8$), over 4 square degrees. The shear correlation functions are
measured on scales from $1'$ to $15'$, and are used to constrain cosmological
parameters. We ensure that our measurements are free from instrumental
systematics \com{by performing a series of tests, including a decomposition of
the signal into $E$- and $B$-modes. We also reanalyse the data independently,
using the shear measurement pipeline developed for the COMBO-17 survey. This
confirms our results and also highlights various effects introduced by
different implementations of the basic ``KSB'' shear measurement method.} We
find that the normalisation of the matter power spectrum on 8 $h^{-1}$Mpc
scales is $\sigma_8=(1.02\pm 0.15)(0.3/\Omega_m)^{0.5}$, where the 68\%CL error
includes noise, sample variance, covariance between angular scales, systematic
effects, redshift uncertainty and marginalisation over other parameters. We
compare these results with other cosmic shear surveys and with recent
constraints from the WMAP experiment. \end{abstract}


\section{Introduction} \label{intro}

Weak gravitational lensing by large-scale structure, or ``cosmic shear'', has
emerged as a powerful cosmological probe, as it is directly sensitive to
foreground mass \citep[for reviews, see][]{bs,bern99,melrev,refrev,witrev}. A
measurement of cosmic shear is therefore closely tied to cosmological theories,
which are principally concerned with the distribution of dark matter. In
particular, the systematic biases of this technique are not limited by unknown
physics such as biasing \citep{oferbias,meg901,hoebias,smithbias,weinbias} or
the mass-temperature relation for X-ray selected galaxy clusters
\citep{hutw,pier,vlnew}. 

Cosmic shear surveys are rapidly growing in size and precision
\citep{bmer,browncs,hamanacs,hoecs,jarviscs,ref02,rrgmeas,vw02}. Cosmological
parameter constraints from these surveys are now approaching the precision of
other methods.

However, cosmic shear surveys can be subject to several systematic biases of
their own. Imperfect telescope tracking, telescope flexure or optical
misalignment within the camera, even at a level that is acceptable for most
purposes, can artificially distort images and align the shapes of distant
galaxies in a way that mimics cosmic shear. 

The survey described in this paper represents a culmination of effort at the
William Herschel Telescope. We have combined the experience of instrumentalists
with detailed image simulations and careful data analysis to control the
various sources of systematic error. Our first cosmic shear paper \citep{bre}
reported an initial detection of cosmic shear using a 0.5 square degree survey
with the William Herschel Telescope (WHT). The second paper \citep{bmer}
compared the WHT shear signal with an independent measurement using the Keck~II
telescope, and examined systematics from these two very different instruments.
In this paper, we extend our WHT survey to cover 4 square degrees to constrain
cosmological parameters, while paying great care in monitoring and correcting
systematic effects. \com{We also now test our entire pipeline for shear
measurement and cosmological parameter estimation by comparing it to external
code, developed independently for the COMBO-17 survey by \citet{browncs}.}

This paper is organised as follows.  In \S\ref{observations} we describe our
survey strategy and observational parameters. In \S\ref{results} we present our
results and draw constraints upon cosmological parameters. In
\S\ref{systematics} we test for the absence of any systematic errors. \com{In
\S\ref{mlb}, we present a second set of results obtined via an independent
shear measurement pipeline}. We conclude in \S\ref{conclusions}.

\section{Observations and Data Reduction}  \label{observations}

\begin{figure}
\centering
\epsfig{figure=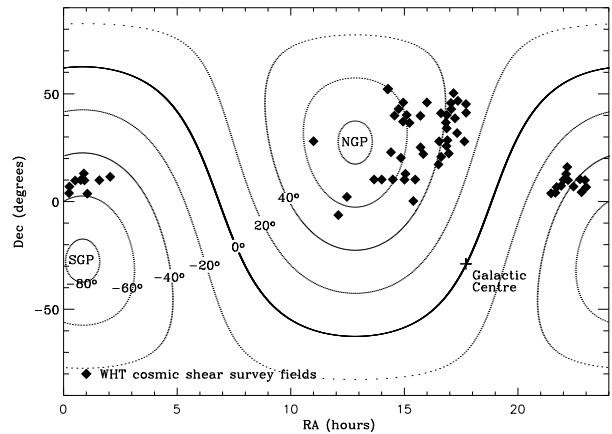,width=80mm}
\caption{The locations of WHT pointings in our deep
pencil-beam survey strategy. The galactic latitudes were tuned to
provide sufficient foreground stars within each image to successfully
model and correct for variations in the PSF.}
\label{fig:skymap}
\end{figure}

We have acquired 4 square degrees of imaging to $R=25.8$ (for a point
source at $5\sigma$) with the {\em Prime Focus Imaging Camera} (PFIC)
of the William Herschel Telescope on La Palma. The median seeing was
$0.69\arcsec$ and no exposures had seeing worse than $1\arcsec$. The
pixel size is $0.24\arcsec$. As shown in figure~\ref{fig:skymap},
pointings were scattered randomly in a pencil-beam survey between
galactic latitudes of $30\degr$ and $70\degr$. This was tuned to
provide $\sim1.5$ stars per arcmin$^2$, with which we could measure
the Point Spread Function (PSF) across each field. The only selection
criterion was to avoid foreground stars brighter than $R\approx 11$ in
the Digitised Sky Survey or APM (Automated Plate Measuring machine)
catalogues.

Cosmic shear statistics have already been presented from the first square
degree of this survey in \citet[][]{bmer}. That data consisted of eight
$8\arcmin\times 16\arcmin$ images and eleven $16\arcmin\times 16\arcmin$ images
taken after the addition of a second, identical CCD to the PFIC. During June
and August 2002, \com{we observed an additional 41 $16\arcmin\times 16\arcmin$
pointings. When combining statistical measurements from all of these fields, we
weight the contribution of the large fields to be twice that of the small
fields. This scheme is ideal for measurements on small angular scales, where
the large fields do indeed contain twice as many pairs of galaxies as the small
fields. The scheme is non-optimal in exploiting the extra signal on large
scales contained via additional pairs across different CCDs. However, in
general, as both large and small fields were randomly deployed, no bias should
result from our weighting scheme.}

For each survey field we took four 900s exposures, each dithered by a few
arcseconds from the last. This strategy enabled a continual monitoring of
astrometric distortions within the telescope, cosmic ray removal, and lower
overheads in the event of inclement weather. Data reduction then proceeded for
the exposures exactly as in \citet{bmer}: \com{see that paper for more
details}. After bias subtraction and flat fielding, fringing remained in the
$R$-band images. This could have been prevented by observing at a shorter
wavelength, but at a cost to the observed number density of background sources.
To remove this, a fringe frame was compiled from all the exposures in each
night. A multiple of this was subtracted from each image which minimised
fringing, to a negligible level $<0.05$\% of the background noise. The four
dithers for each field were then re-aligned (using linear interpolation between
adjacent pixels to allow sub-pixel offsets) and stacked (with
$3\sigma$-clipping to remove cosmic rays). 

Objects were located on the final images using {\sc hfindpeaks} from the {\sc
imcat} package by \citet[][KSB]{ksb}. Following the recommendations of
\citet{kecksys}, objects within 10$\arcsec$ of saturated stars or 5$\arcsec$ of
the edge of the CCDs were masked and removed. We also removed noisy objects
from the catalogue with cuts in size, signal-to-noise and ellipticity of
$r_g>1\rmn{~pixel}$, $\nu>15$ and $|\varepsilon|<0.5$. \com{This is the same
procedure to that adopted in \citet{bmer}, and leaves 15.2 galaxies per
arcmin$^{2}$ in the final catalogue}, with a median magnitude of $23.5\pm 0.2$.
According to \citet{cohen} \com{as before}, this corresponds to a median source
redshift of $z_s\simeq 0.80\pm 0.06$. 

\com{Galaxies' observed ellipticities, $\varepsilon$, were formed from
combinations of their Gaussian-weighted quadrupole moments, then corrected for
convolution with the telescope's PSF. The shape of the PSF was measured from
stars in each image, then interpolated to the positions of galaxies via a
polynomial fit within each CCD separately. The order of this polynomial was
varied from field to field: it was occasionally raised to fourth order but
lowered to second order where possible. This limited spurious power on small
scales and particularly around the edges of the field. Even so, an acceptable
polynomial fit to the PSF was not always possible. We conservatively discarded
14 fields, all of which were otherwise within specifications, but whose stellar
ellipticities remained correlated after correction. More sophisticated PSF
fitting methods have recently been suggested by \citet{henkpsf} and
\citet{jarvispca}. Such techniques may improve the interpolation, and thus
increase the number of useable fields, but they have not been investigated
here. 

The shear susceptibility factor for each galaxy, $P^\gamma$, was determined
from the higher-order shape moments of the galaxy and the interpolated PSF. The
shear susceptibility is a notoriously noisy quantity; in this implementation of
KSB, we chose to fit $P^\gamma$ with a cubic polynomial as a function of galaxy
size, $r_g$. We can then form shear estimators for each galaxy,
$\gamma=\varepsilon/P^\gamma$. We apply a final calibration factor of $(0.85\pm
0.04)^{-1}$ to these shears. This calibration factor was determined by
\citet{bacsim}, using our shear measurement pipeline upon simulated WHT images
with a known input signal.}

\begin{figure}
\centering
\epsfig{figure=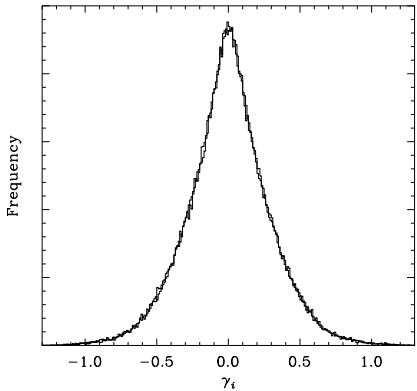,width=55mm}
\caption{The distribution of measured shear estimators from all galaxies used
in the WHT survey. A fitting function is given in the text. The two components
of shear are shown superimposed; their distibutions are indistinguishable.
Notice, however, the extended wings typical of the KSB method.}
\label{fig:gdist}
\end{figure}

Each set of four dithered exposures were also used to continually monitor
astrometric distortions within the telescope. As observed in \citet{bacsim},
these closely follow the engineering predictions in the PFIC manual of
$\gamma_\rmn{tangential}=0$, $\gamma_\rmn{radial}=-8.2\times 10^{-5}r^2$ with
$r$ measured in arcminutes from the field centre. \com{This affects the shear
estimate of an average galaxy in a large exposure by less than $2\times
10^{-3}$. A galaxy in the corner of a large exposure is altered by $1\times
10^{-2}$. Since this is no longer an {\it entirely} negligible effect in our
enlarged survey, we subtract this distortion from the final shear catalogues
using the shear addition and subtraction operators in \citet{ssss}.  

The resulting distribution of shear estimators is shown in
figure~\ref{fig:gdist}. Both components of shear are well-fit by the normalised
proabability density function
\begin{equation}
P(\gamma_i)\approx 0.26 \frac{\exp(-x^2/0.65^2)}{(x^2+0.27^2)} ~.
\label{eq:gdist}
\end{equation}
Extended, Cauchy-like wings are a salient feature of the KSB method, where
ellipticity measurements are formed from ratios of (noisy) quadrupole moments.}

\section{Results} \label{results}

\subsection{Shear-shear correlation functions} \label{cth_gg}

The power spectrum of the weak lensing shear is given by
\begin{equation}
C^\gamma_{\ell} ~=~ \frac{9}{16} \left( \frac{H_{0}}{c} \right)^{4} \Omega_{m}^{2}
  \int_{0}^{\chi_h} \left[ \frac{g(\chi)}{D_A(\chi)} \right]^{2}
  P\left(\frac{\ell}{r}, \chi\right)~\rmd\chi,
\label{eq:shearpowerspectrum}
\end{equation}
where $\chi$ is a comoving distance; $\chi_h$ is the horizon distance;
$D_A(\chi)$ is an angular diameter distance; $g(\chi)$ is the lensing
weight function; and $P(k,z)$ is the underlying 3D distribution of
mass in the universe. The \com{two-point} shear correlations functions can 
be expressed in terms of the power spectrum as
\begin{eqnarray}
C_1(\theta) & = &
\frac{1}{4\pi}\int_0^\infty C_\ell^\gamma ~\big[{\rm J}_0(\ell\theta)
+{\rm J}_4(\ell\theta)\big] ~\ell~\rmd\ell 
\label{eq:c1} \\
C_2(\theta) & = &
\frac{1}{4\pi}\int_0^\infty C_\ell^\gamma ~\big[{\rm J}_0(\ell\theta)
-{\rm J}_4(\ell\theta)\big] ~\ell~\rmd\ell ~.
\label{eq:c2}
\end{eqnarray}

These can be measured by averaging over galaxy pairs, as
\begin{eqnarray}
C_{1}(\theta) & = & 
  \langle ~ \gamma_{1}^r(\bmath{r}) ~
            \gamma_{1}^r(\bmath{r}+\bmath{\theta}) ~
            \rangle
\label{eq:cth_c1} \\ 
C_{2}(\theta) & = &
  \langle ~ \gamma_{2}^r(\bmath{r}) ~
            \gamma_{2}^r(\bmath{r}+\bmath{\theta}) ~
            \rangle ~,
\label{eq:cth_c2}
\end{eqnarray}
where $\theta$ is the separation between the galaxies and the
superscript~$^r$ denotes components of shear rotated so that $\gamma_1^r$
($\gamma_2^r$) in the first galaxy points along (at 45$\degr$ from) the vector
between the pair. A third shear-shear correlation function can be formed,
\begin{equation}
C_{3}(\theta) ~=~
  \langle ~ \gamma_{1}^r(\bmath{r}) ~
            \gamma_{2}^r(\bmath{r}+\bmath{\theta}) ~
            \rangle ~+~
  \langle ~ \gamma_{2}^r(\bmath{r}) ~
            \gamma_{1}^r(\bmath{r}+\bmath{\theta}) ~
            \rangle ~,
\label{eq:cth_c3}
\end{equation}
for which the parity invariance of the universe requires a zero
signal. $C_3(\theta)$ can therefore be used as a first test for
the presence of systematic errors in our measurement.

\begin{figure}
\centering
\epsfig{figure=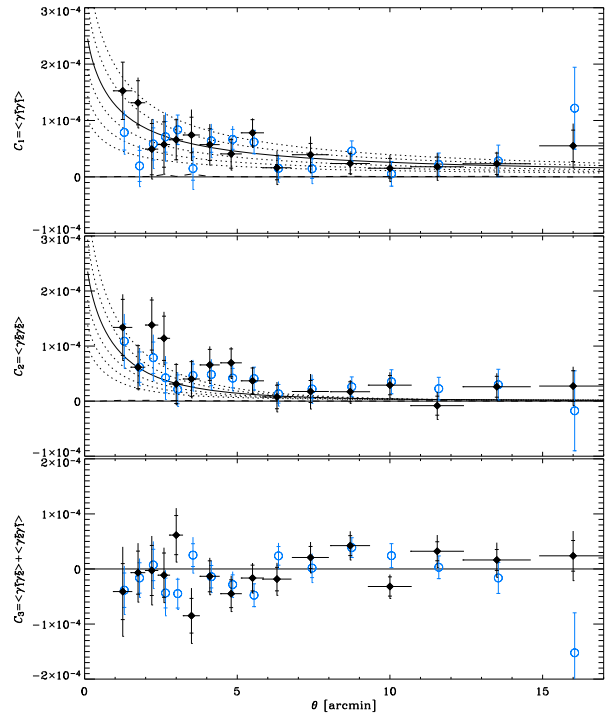,width=80mm}
\caption{
Correlation functions of the shear field measured in our 4 square degree WHT
survey. The solid data points show our measurement. The inner error bars are
for statistical errors only; the outer error bars also include full
non-Gaussian sample variance. 
The blue circles show measurements of our data by the COMBO-17 pipeline,
artificially adjusted to correct for its slightly higher source
redshift distribution and permit easy comparison. 
The solid line shows the theoretical prediction for a $\Lambda$CDM model with
$\Omega_m=0.3$, $\Omega_\Lambda$=0.7, $\Gamma=0.21$ and $\sigma_8=1.0$,
assuming a median source redshift of $z_s=0.8$ and using the fitting functions
of \citet{smith}. The dotted lines show similar theoretical predictions, but
with $\sigma_8$ ranging from 0.7 (bottom) to 1.2 (top). The dashed lines barely
visible above the $x$~axis show
the correlation of galaxy shears with the observed PSF anisotropy.}
\label{fig:cth}
\end{figure}

To perform the measurement in practice, we first measure the shear correlation
functions in each field, using all galaxy pairs within various $\theta$ bins.
To obtain a combined result for the entire survey, we then average the binned
values for each field. \com{We find that the measured correlation functions are
quite sensitive to changes in the binning scheme. The highly non-Gaussian
distribution of shear estimators (see figure~\ref{fig:gdist}), plus additional
non-Gaussianity introduced by computing correlation functions, means that the
inclusion in one bin of a small number of outlying fields can significantly
bias measurements. Both the correlation functions and the subsequent
constraints on cosmological parameter constraints move within their full
1$\sigma$ statistical error bars.} During the averaging, we therefore introduce
$3\sigma$-clipping in each bin to remove some outliers, and further
$3\sigma$-clipping in $C_3(\theta)$ and the star-galaxy cross-correlation
functions $C^{\mathrm SG}_1(\theta) + C^{\mathrm SG}_2(\theta)$ to eliminate
flaws with PSF correction (see \S\ref{psfcorrect}). \com{This leaves between
$N_f=40$ and $N_f=43$ fields used for each angular bin. Our final choice of bin
size yields representative central values. The result is shown in
figure~\ref{fig:cth}. We shall separately estimate the amount of instability
caused by binning, and increase the error upon cosmological parameter
constraints accordingly.}

\begin{figure}
\centering
\epsfig{figure=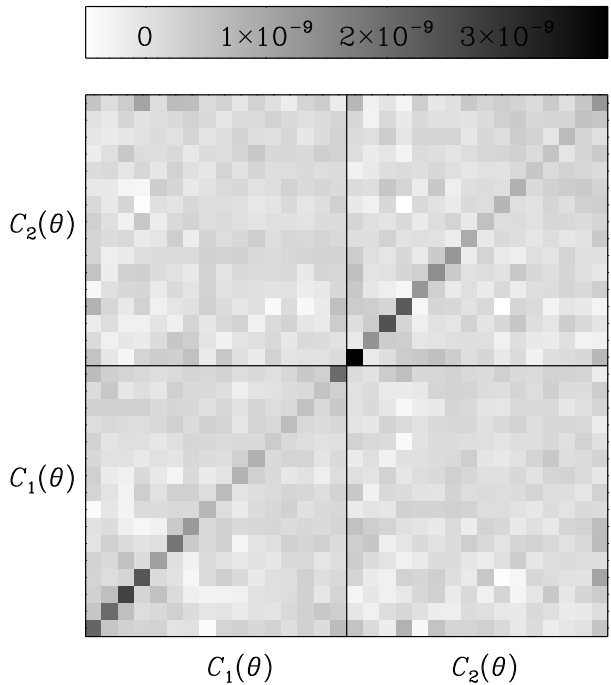,width=80mm}
\caption{\com{Covariance matrix of different angular bins of the shear 
correlation functions shown in
figure~\ref{fig:cth}, from small to
large $\theta$, then from small to large $\theta$ again. For example,
the bottom-left quarter shows ${\rm cov}[C_{1}(\theta),C_{1}(\vartheta)]$.
The diagonal elements within this are the variances in each bin.}}
\label{fig:cov}
\end{figure}

In order to derive any constraints on cosmological parameters, it will
also be necessary to know the covariance of $C_i(\theta)$ between
different angular bins. Our pencil-beam survey strategy with many
independent fields makes it easy to measure their covariance matrix
\begin{eqnarray}
{\rm cov}[C_{i}(\theta),C_{j}(\vartheta)] ~ \simeq 
~~~~~~~~~~~~~~~~~~~~~~~~~~~~~~~~~~~~~~~~~~~\nonumber \\
\frac{1}{N_f^{2}} 
\sum_{f=1}^{N_f} \left[ C_{i}^{f}(\theta) - C_{j}(\theta) \right] 
\left[ C_{i}^{f}(\vartheta) - C_{j}(\vartheta) \right] ~,
\label{eq:covariance}
\end{eqnarray}
\noindent where the summation is over all fields, and the
superscript~$^f$ denotes correlation functions calculated in one field
alone. This matrix is depicted in figure~\ref{fig:cov}, and shows the
significant covariance, especially between adjacent bins.

\subsection{Cosmological parameter constraints} \label{constraints}

We now use a Maximum Likelihood method to determine the constraints set by our
observations upon the cosmological parameters $\Omega_m$, the total
mass-density of the universe, and $\sigma_8$, the normalisation of the matter
power spectrum at 8 $h^{-1}$ Mpc. The analysis directly uses the observed
correlation functions $C_1(\theta)$ and $C_2(\theta)$, proceeding as in
\citet{bmer}, except that theoretical predictions for the non-linear power
spectrum are calculated via the updated fitting functions of \citet{smith}
rather than those by \citet{pandd}. This has the effect of lowering our final
constraint on $\sigma_8\Omega_m^{0.5}$ by about $5$\%. \com{We use the fitting
functions for the linear transfer function suggested by \citep{bbks}.} Note
that, although we shall perform an $E/B$ decomposition, we fit $C_1$ and $C_2$
rather than the $E$ mode signal to avoid degeneracies arising from our survey's
finite size. We will use the $E/B$ decomposition as an a posteriori consistency
check for systematics on relevant scales; because of such contamination, we
discarded the first and last data points shown in figure~\ref{fig:cth} (see
discussion in \S\ref{systematics}).

The theoretical correlation functions were first calculated from
equation~(\ref{eq:shearpowerspectrum}) on a 2D grid across the $\Omega_m$ {\it
vs} $\sigma_8$ plane. \com{The power spectrum shape parameter was set to
$\Gamma=0.21$,} consistent with recent observations of clustering in galaxy
redshift surveys \citep{percival,szalay}. The median redshift for source
galaxies was fixed to $z_s=0.8$ for WHT and $z_s=1.0$ for Keck. Errors on these
parameters will be propagated separately into our final constraints.

\begin{figure}
\centering
\epsfig{figure=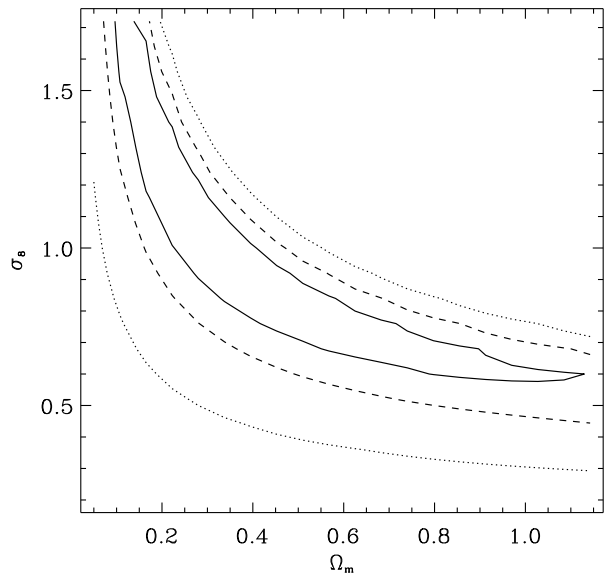,width=80mm}
\caption{Constraints upon cosmological parameters $\Omega_m$ and
$\sigma_8$, from a maximum-likelihood analysis of our WHT cosmic shear
survey data. The 68.3\% (solid), 95.4\% (dashed) and 99.7\% (dotted)
confidence limits include statistical errors and non-Gaussian cosmic
variance. However, they include neither the calibration of the shear
measurement method, nor uncertainty in the source galaxy redshift
distribution. These sources of error are considered separately in the
text.}
\label{fig:wht_constraints}
\end{figure}

\begin{figure}
\centering
\epsfig{figure=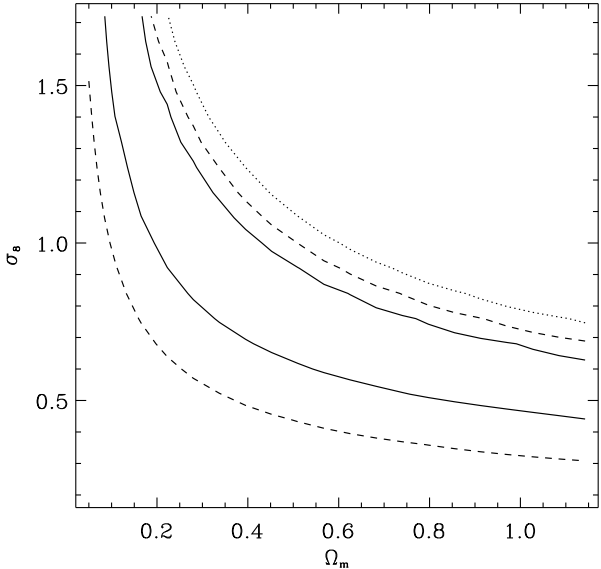,width=80mm}
\caption{ Constraints upon cosmological parameters from the Keck
cosmic shear survey by \citet{bmer}, showing the 68.3\%, 95.4\% and 99.7\%
confidence limits as in figure~\ref{fig:wht_constraints}. Only one
edge of the 99.7\% confidence contour is visible inside this parameter
range.}
\label{fig:keck_constraints}
\end{figure}

We then fitted the observed shear correlation functions $\vec{d}(\theta)$ to
the theoretical predictions calculated at the centres of each bin
$\vec{t}(\theta)$, computing the log-likelihood function
\begin{eqnarray}
\chi^2 & = & \big(\vec{d}(\theta) - \vec{t}(\theta,\Omega_m,\sigma_8) \big)^T \nonumber \\
 ~ & ~ & ~~~~~\Big({\rm cov}[C_{i}(\theta),C_{j}(\vartheta)]\Big)^{-1} ~
\big(\vec{d}(\vartheta) - \vec{t}(\vartheta,\Omega_m,\sigma_8) \big)
\label{eq:chisq}
\end{eqnarray}
throughout the grid. We thus explore parameter space in this plane,
and minimise $\chi^2$ to find the best-fit cosmological model. To compute
confidence contours, we numerically integrate the likelihood function
\begin{equation}
L(\Omega_m,\sigma_8) ~=~ e^{-\chi^2/2} ~.
\end{equation}

Note that the matrix inversion in equation~(\ref{eq:chisq}) is unstable,
possibly for the same reasons as the binning instability. Since ${\rm
cov}[C_{1}(\theta),C_{2}(\theta)]$ is constitent with zero but noisy, we solve
this problem in practice by forcing it to zero. The inversion is then
well-behaved.

Our constraints on cosmlogical parameters are presented in
figure~\ref{fig:wht_constraints}, and the constraints from our Keck survey
\citep{bmer} are reproduced in the same format in
figure~\ref{fig:keck_constraints}. In both cases, the contours show 68.3\%,
95.4\% and 99.7\% confidence limits, including statistical errors and
non-Gaussian sample variance. They reveal the well-known degeneracy between
$\Omega_m$ and $\sigma_8$ when using only two-point statistics.

\begin{figure}
\centering
\epsfig{figure=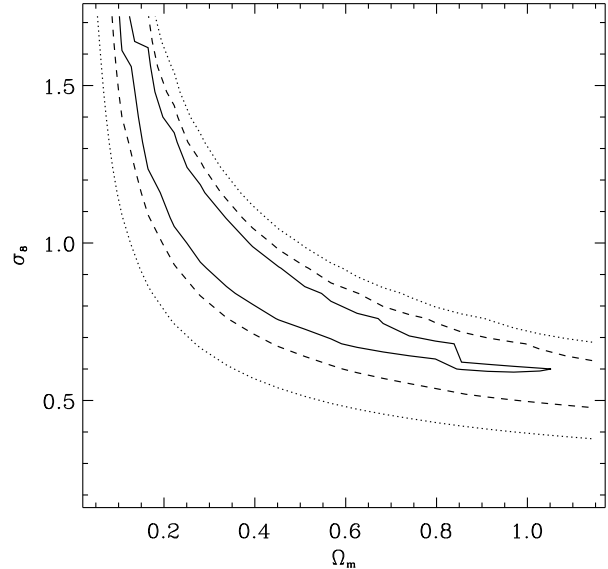,width=80mm}
\caption{ Constraints upon cosmological parameters for the combination of
both WHT and Keck surveys, showing the 68.3\%, 95.4\% and 99.7\%
confidence limits, as in figure~\ref{fig:wht_constraints}.}
\label{fig:combined_constraints}
\end{figure}

A good fit to the 68.3\% confidence level from our WHT data is given by
\begin{equation}
\sigma_{8} \left( \frac{\Omega_{m}}{0.3} \right)^{0.50} = 1.02 \pm 0.13 ~,
\label{eq:wht_con}
\end{equation}
while the Keck data is well-fit by 
\begin{equation}
\sigma_{8} \left( \frac{\Omega_{m}}{0.3} \right)^{0.52} = 1.01 \pm 0.19 ~,
\label{eq:keck_con}
\end{equation}
The multiplication of the respective likelihood functions provides a
constraint from a combined survey. Such confidence contours are shown
in figure~\ref{fig:combined_constraints}, with the 68.3\% confidence
level well-fit by
\begin{equation}
\sigma_{8} \left( \frac{\Omega_{m}}{0.3} \right)^{0.50} = 1.02 \pm 0.12 ~,
\label{eq:combined_con}
\end{equation}
for $0.1<\Omega_m <0.7$. 

Note that all of these constraints include only the statistical error
and sample variance. We can propagate other sources of error by noting
that
\begin{equation}
C_i(5')\propto \Omega_m^{1.46}\sigma_8^{2.45}z_s^{1.65}\Gamma^{-0.11}(P^\gamma)^{-2},
\label{eq:coszgp}
\end{equation}
where $i=1$ and $2$ and $P^\gamma$ is the shear calibration factor, in
a fiducial $\Lambda$CDM cosmological model with $\Omega_m=0.3$,
$\Omega_\Lambda$=0.7, $\Gamma=0.21$ and $\sigma_8=1.0$. Adding in turn to our
constraint~(\ref{eq:combined_con}): a 6\% additional binning instability, 
a 10\% source redshift uncertainty, a 15\% prior on $\Gamma$, and a 5\% shear
calibration uncertainty, gives a final 68.3\% CL constraint for the combined 
survey of
\begin{eqnarray}
\sigma_{8} \left(
\frac{\Omega_{m}}{0.3} \right)^{0.5} = 1.02 \pm 0.12 \pm 0.06 \pm 0.066 \pm 0.006 \pm 0.04
\label{eq:con_full_errors} \\
  = 1.09 \pm 0.15 ~,~~~~~~~~~~~~~~~~~~~~~~~~~~~~~~~~~~~~~~~~~~~~~~~~~~~~~
\label{eq:con_full_error}
\end{eqnarray}
where the various errors have been combined in quadrature on the second line.
This result now includes all contributions to the total error budget:
(non-Gaussian) statistical noise and sample variance, covariance between
different angular scales, binning instability, shear calibration error, source
redshift uncertainty, and marginalisation over $\Gamma$.

\subsection{Shear variance}

For historical reasons, cosmic shear results are often expressed as the
variance of the shear field in circular cells on the sky. For a top-hat cell
of radius $\theta$, this measure is related to the shear correlation
functions by
\begin{eqnarray}
\sigma_\gamma^2 & \equiv & \langle ~ |\overline{\gamma}|^2 ~ \rangle
                ~~ = ~ \frac{2}{\pi}\int_0^\infty
                C_\ell^{\gamma}(\ell) ~\big[{\rm
                J}_1(\ell\theta)\big]^2~\ell~\rmd\ell
\label{eq:shvar} \\
~ & \simeq & \frac{2}{\theta^2}
\int_0^\theta \left[ C_1(\vartheta)+C_2(\vartheta) \right]~
     \rmn{d}\vartheta ~,
\end{eqnarray}
where we have used a small angle approximation involving the Bessel
functions.  Note that the shear variance is more strongly correlated
on different angular scales in this form than they are as correlation
functions.

\begin{figure}
\centering
\epsfig{figure=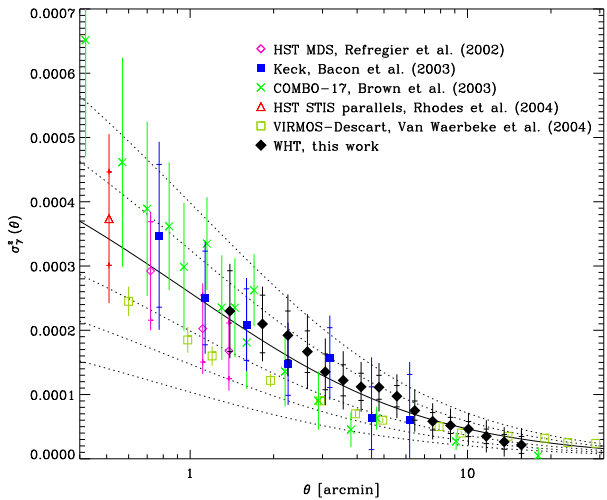,width=80mm}
\caption{Shear variance in (circular, top-hat) cells, as a function of
the radius of the cells. Our results are compared to those from
similarly deep surveys by other groups. \com{
For ease of comparison, all results have been rescaled as if their median
source redshift had been the same (see text). 
As in figure~\ref{fig:cth}, the theoretical curves are predictions for a
$\Lambda$CDM cosmology, with $\Omega_m=0.3$, $\Omega_\Lambda=0.7$, 
$\Gamma=0.21$ and $\sigma_8$ ranging from 0.7 (bottom) to 1.2 (top).}}
\label{fig:var}
\end{figure}

In practice, data is not available on all the scales necessary to perform this
integration. The correlation functions have not been calculated on scales
smaller than 1$\arcmin$, and are contaminated by systematics on scales smaller
than 2$\arcmin$ (see \S\ref{systematics}). We determine the deficit in the
measured values as a function of $\theta$ by extrapolating the data through
these scales using the theoretical predictions given by the best-fit
cosmological model determined in \S\ref{constraints}. This deficit ($\sim
2\times10^{-5}$ at $3\arcmin$ and $1\times10^{-5}$ at $5\arcmin$) is then
added back on to our measured data points. 

We present our results as the variance of shear in cells, and compare them to
those from similarly deep lensing surveys in figure~\ref{fig:var}. These
surveys use data from 8.5 square degree, $z_s=0.8$ VIRMOS-Descart survey on the
3.6m CFHT by \citet{vw04}; the 1.25 square degree, $z_s=0.85$ COMBO-17 survey
on the 2.2m La Silla telescope by \citet{browncs}; the 0.36 square degree,
$z_s=0.9$ Medium Deep Survey (MDS) with the {\em Wide Field and Planetary
Camera} on HST by \citet{ref02}; 0.27 square degrees of random fields observed
to $z_s=1.0$ in parallel-mode with the {\em Space Telescope Imaging
Spectrograph} (STIS) on HST by \citet{jr_stis}; and the 0.6 square degree,
$z_s=1.0$ pencil-beam survey using the 10m Keck~II telescope by \citet{bmer}.
\com{For ease of comparison between these different surveys, all the results
have been scaled to the values that would have been obtained if their median
source redshift had been $z_s=0.8$, and using the theoretical prediction that
$\sigma_\gamma^2\propto z_s^{1.65}$. Results from deeper surveys are thus shown
slightly lower here than in their original papers.}

\section{Tests for systematic Biases} \label{systematics}

The validity of any cosmic shear result depends sensitively upon the treatment
of systematic errors and the control of observational biases. Almost all
systematic effects, whether they be due to a background gradient, astrometric
distortions within the telescope or imperfectly corrected PSF anisotropy, act
to increase the observed correlations between galaxy shapes. All the effects
can mimic cosmic shear and the most important task incumbent upon any weak
lensing survey is to prove that its systematics are controlled to a negligible
level. As already described in \S\ref{observations}, the astrometric
distortions in WHT have been corrected for, and the basic data reduction was
performed sufficiently carefully to eliminate most biases. In this section, we
discuss further tests for other sources of residual systematics.

\subsection{E-B decomposition} \label{eb}

The correlation functions can be recast in terms of $E$ (gradient) and $B$
(curl) modes of the shear field \citep{crit00,peneb}. Gravitational lensing is
expected to produce only $E$ modes, except for a very low level of $B$ modes
due to lens-lens coupling along a line of sight \citep{schneb}.  Systematic
effects are likely to affect both $E$- and $B$-modes equally. The presence of
any non-zero $B$ mode would therefore be a useful indication of contamination
from other sources.

$E$- and $B$-modes correspond to patterns within an extended region on the sky.
The separation cannot be performed locally, but requires the shear correlation
functions to be integrated over a wide range of angular scales. In practice, we
cannot perform these integrals exactly because our correlation function data
extends only between $\sim 2\arcmin$ and $16\arcmin$. In other words, a shear
field measured within a finite aperture can not be uniquely split into distinct
$E$- and $B$-mode components: some correction will always be necessary. 

\com{We find the frequently-used decomposition into aperture mass $M_{\rm
ap}(\theta)$ and $M_\perp(\theta)$ modes \citep{bs} to be particularly unstable
with our data. The correlation functions need to be integrated between 0 and
$2\theta$, with a lot of weight placed upon their values at small angular
scales. Since these are changing rapidly, the end result becomes even more
sensitive to the bin spacing. Furthermore, our measured correlation functions
are least reliable at small separations. This is likely to be the case in any
real data because of small-scale effects like overlapping galaxy isophotes.
Since data from these small scales need to be included in all subsequent
integrals, any bias there would adversely affect the aperture mass on all
scales.}

\begin{figure}
\centering
\epsfig{figure=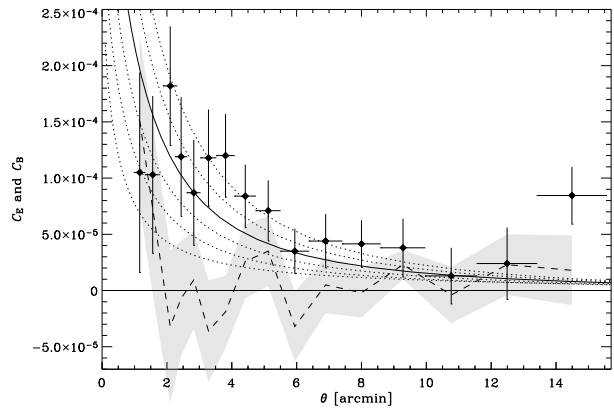,width=80mm}
\caption{$E$-$B$ decomposition of the shear field observed in our WHT survey.
The points show the measured $E$ (tangential) modes of the shear field. As in
figure~\ref{fig:cth}, the theoretical curves are predictions for a $\Lambda$CDM
cosmology, with $\Omega_m=0.3$, $\Omega_\Lambda=0.7$,  $\Gamma=0.21$ and
$\sigma_8$ ranging from 0.7 (bottom) to 1.2 (top). The dashed line line shows
our measured $B$-mode (curl) signal, and the shaded region shows its 1$\sigma$
error bar. In the absence of systematics, the $B$-mode should be consistent
with zero.} 
\label{fig:eb} 
\end{figure}

We therefore prefer another method for $E/B$ decomposition. Following
\citet{crit00}, we calculate
\begin{equation}
C_E(\theta) \equiv
           C_1(\theta)+2\int_\theta^\infty
           \left( 1-\frac{3\theta^2}{\vartheta^2} \right)
           \frac{C_{1}(\vartheta)-C_{2}(\vartheta)
           }{\vartheta}~\rmn{d}\vartheta ~,
\end{equation}
which contains only the $E$-mode signal and
\begin{equation}
C_B(\theta) \equiv
           C_2(\theta)-2\int_\theta^\infty
           \left( 1-\frac{3\theta^2}{\vartheta^2} \right)
           \frac{C_{1}(\vartheta)-C_{2}(\vartheta)}{\vartheta}~\rmn{d}\vartheta ~,
\label{eq:cth_eb}
\end{equation}
which contains only the $B$-mode signal. These can be calculated using data
exclusively from scales larger than $\theta$. A correction will still need to
be made for absence of data on scales larger than $16\arcmin$, but this is in a
regime where the expected signal (and the necessary correction) is small. As
can be seen from the above equation, a function of $\theta$ (not only a
constant of integration) must be added to $C_E(\theta)$ and subtracted from
$C_B(\theta)$ \citep[c.f.][]{peneb}. We calculate this function by using
theoretical predictions for the best-fit cosmological model (as determined in
\S\ref{constraints}) to extrapolate our data to infinity. The size of this
correction is approximately one third of the size of the measured $E$-mode
signal. The correction is 2.4$\times10^{-5}$ at $5\arcmin$ and
1.5$\times10^{-5}$ at $10\arcmin$. 

An $E$/$B$ decomposition of our data is shown in figure~\ref{fig:eb}. On scales
$1.7'<\theta<15'$, we find a $B$-mode signal consistent with zero, confirming
the absence of systematics on these scales. Note that because of the extra
uncertainty introduced by this additive function, we have not used the derived
$E$-mode signal to fit cosmological parameters. We instead direcly fit the
measurements of $C_1$ and $C_2$ on those scales deemed free of systematic
errors only (see \S\ref{constraints}).

\subsection{PSF correction} \label{psfcorrect}

\begin{figure}
\centering
\epsfig{figure=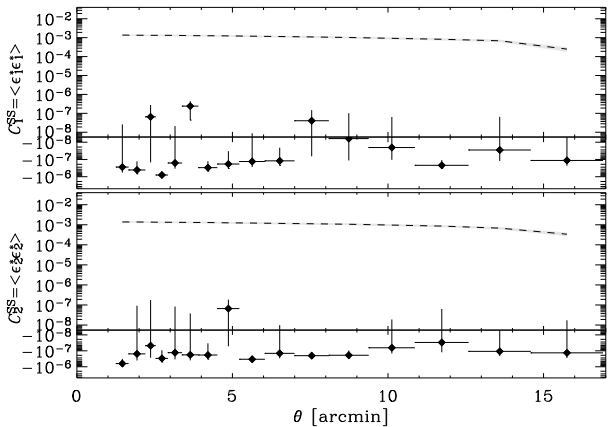,width=80mm}
\caption{Correlation functions of stellar ellipticities before
(dashed line) and after correction (solid points). The shaded region shows the
size of 1$\sigma$ errors on the pre-correction correlation functions.
Application of the KSB method has successfully reduced these by about four 
orders of magnitude.}
\label{fig:cth_stellar}
\end{figure}

The WHT PSF over long exposures can be quite anisotropic, with a mean stellar
ellipticity of $0.051\pm 0.28$, where the error quoted is the rms stellar
ellipticity within one field, averaged over all fields. Application of KSB
reduces this to $0.0056\pm 0.0012$. Figure~\ref{fig:cth_stellar} shows the full
correlation functions
\begin{equation}
C_{i}^{\rm SS} ~ \equiv ~ \langle ~ e^*_i    ~ e^*_i ~ \rangle~ ,
\label{cth_ss}
\end{equation}

\noindent where $i=\{1,2\}$. Application of the KSB method has successfully
reduced these by four orders of magnitude. Note that these are correlations of
ellipticity rather than shear, and therefore exhibit a different overall
normalisation that is connected to the shear susceptibility factor. The
elimination of stellar anisotropy is also a slightly different problem than the
challenging correction of galaxy shapes, where the interpolation of the PSF is
critical, and the size of the weight function $r_g$ may be different. 

The correction of galaxy shapes for the PSF anisotropy can be tested using
cross-correlation functions between corrected shears $\gamma_i$ from
galaxies and stellar ellipticity $e^*_i$ before correction,
\begin{equation}
C_{i}^{\rm SG} ~ \equiv ~ 
  \frac{\langle ~ \gamma_i ~ e^*_i ~ \rangle^2} 
       {\langle ~ e^*_i    ~ e^*_i ~ \rangle~} ,
\label{cth_sg}
\end{equation}

\noindent where $i=\{1,2\}$. These cross-correlation functions are
shown as dashed line in the top two panels of figure~\ref{fig:cth},
and is consistent with zero on all scales. Note that PSF correction
residuals would also have appeared as $B$-modes in
figure~\ref{fig:eb}, which are in fact consistent with zero.

Uncertainties still remain in the KSB method about the overall {\it
calibration} of shear estimators after PSF correction
\citep[e.g.][]{ssss,vw02}. Detailed image simulations by \citet{bacsim} or
\citet{erbsim} have been used to study these issues. \citet{bacsim} found that,
with our survey and telescope parameters, our implementation of KSB requires a
constant calibration factor of $(0.85\pm 0.04)^{-1}$ to be applied to all
calculated shear estimators.

\subsection{Shear as a function of CCD position}

\begin{figure}
\centering
\epsfig{figure=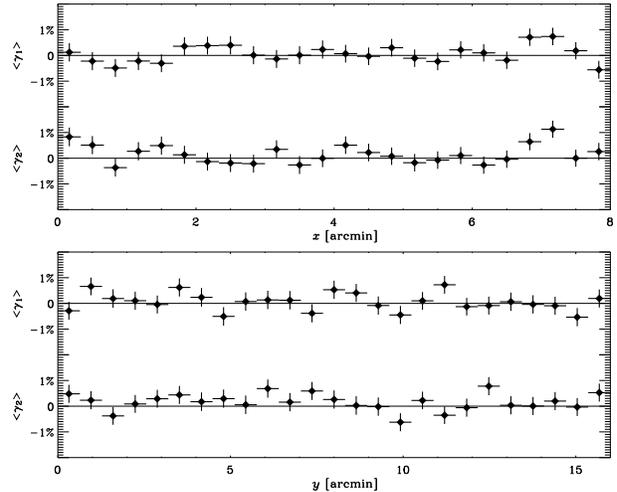,width=80mm}
\caption{Galaxy shears as a function of position on the WHT CCD,
\com{averaged over all the fields used in our survey. That this is consistent
with zero everywhere demonstrates an absence of systematics} concerning CCD
readout, telescope vibration and tracking.}
\label{fig:gxy}
\end{figure}

\begin{figure}
\centering
\epsfig{figure=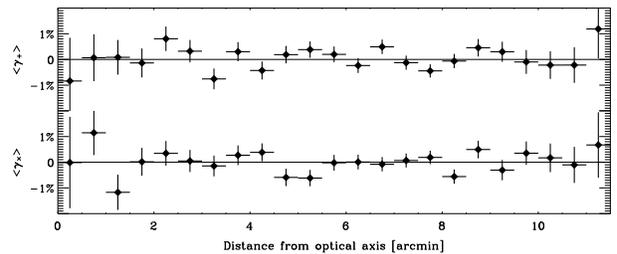,width=80mm}
\caption{\com{The radial and ``curl'' components of galaxy shears as a function
of distance from the field centre/optical axis, averaged over all the fields
used in our survey. That this is conststent with zero on all scales
demonstrates an absence of systematics concerning alt-az tracking, and 
the successful correction of astrometric distortions within the telescope, 
whose expected behaviour is purely radial (see \S\ref{observations}).}}
\label{fig:gr}
\end{figure}

Problems with read noise or charge transfer efficiency on the CCD, or telescope
flexure and vibration, could cause the measured shear to vary as a function of
position on the chip, even when averaged over many separate fields.
Figure~\ref{fig:gxy} shows plots of shear as a function of $x$ and $y$.
Figure~\ref{fig:gr} shows plots of shear as a function of $r$. Both are
consistent with zero, as desired.

\begin{figure*}
\centering
\epsfig{figure=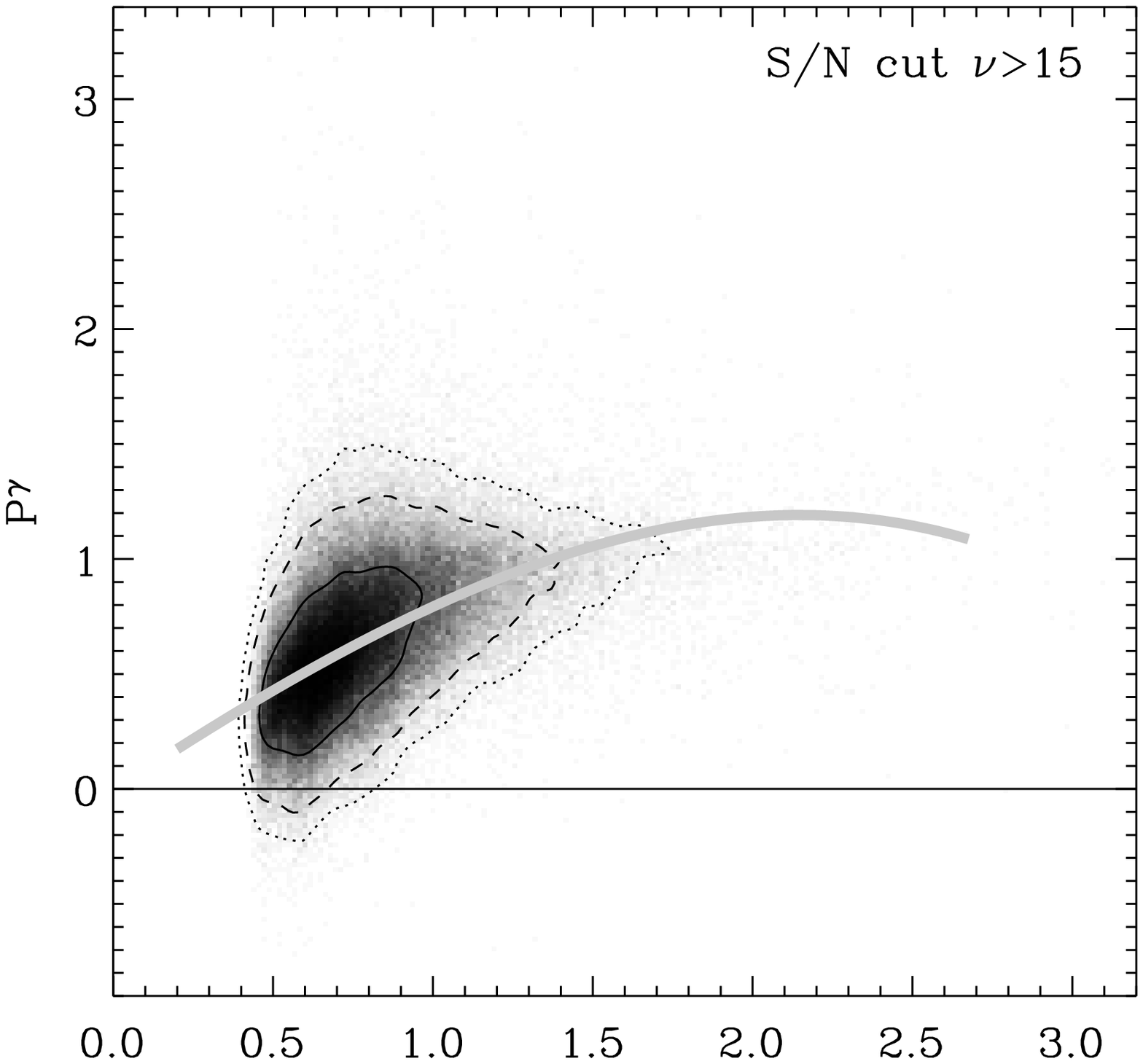,width=58mm,height=55.35mm}~
\epsfig{figure=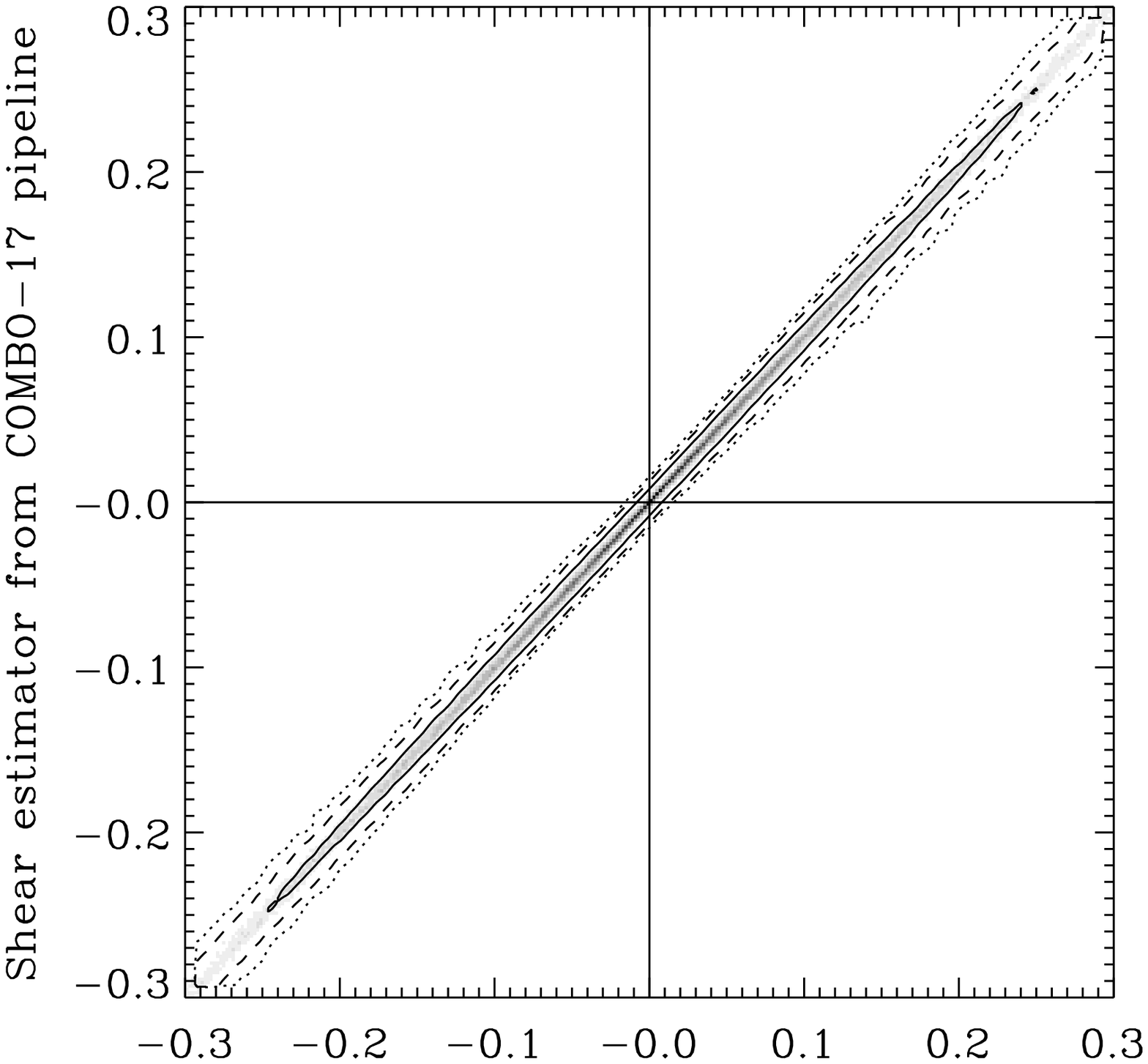,width=58mm,height=55.35mm}
\epsfig{figure=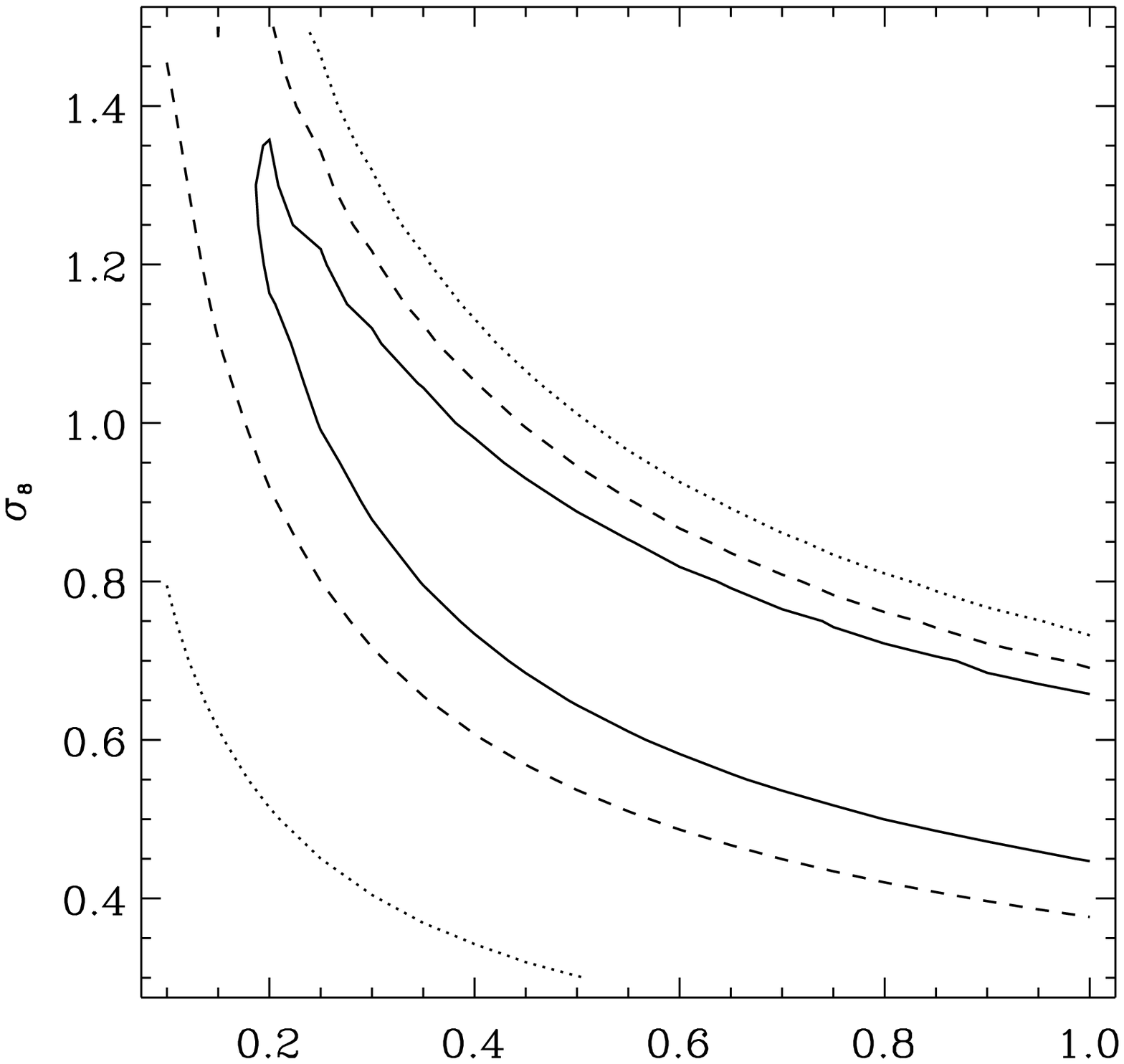,width=58mm,height=55.35mm}
\\
\epsfig{figure=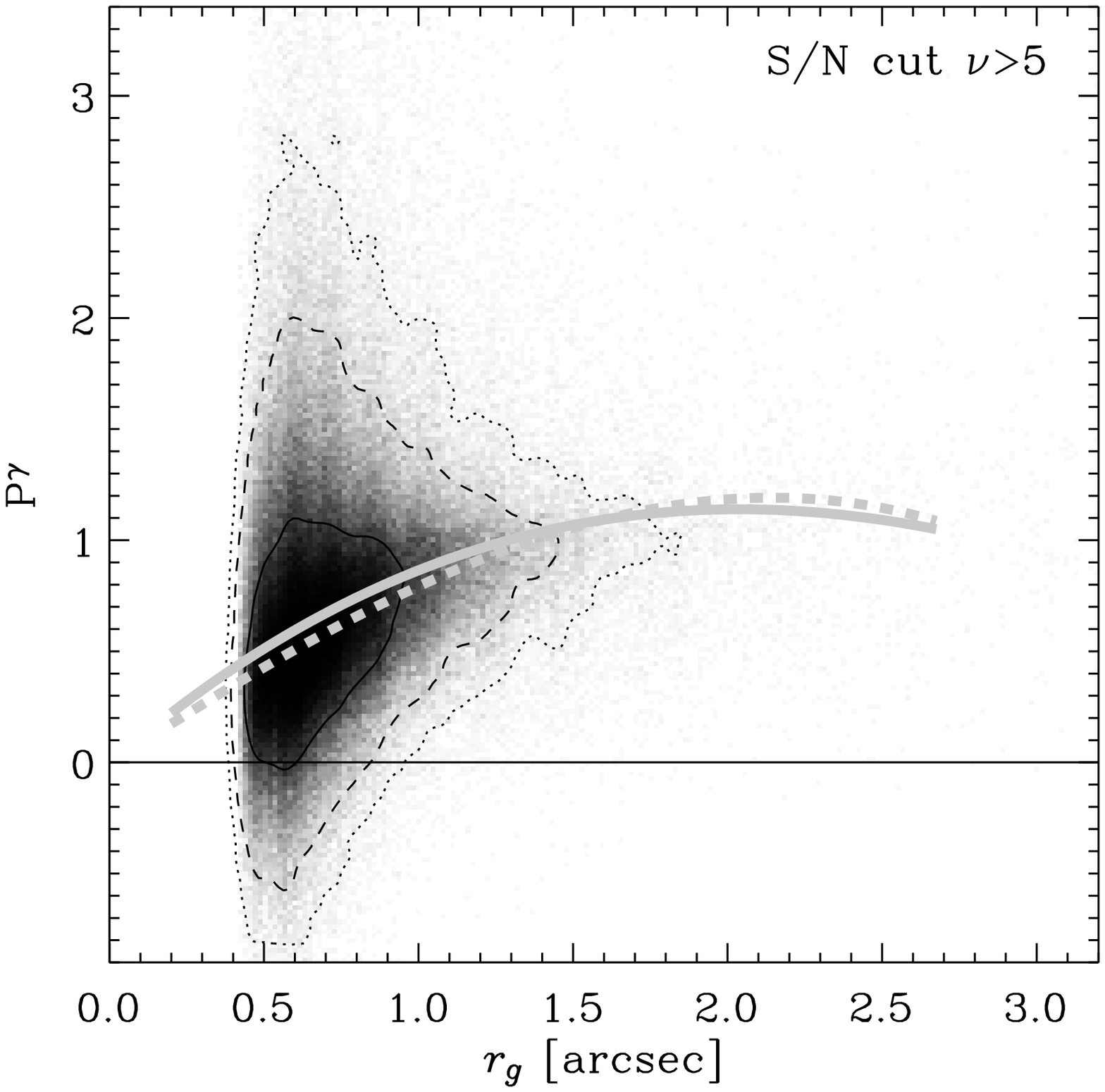,width=58mm,height=58mm}~
\epsfig{figure=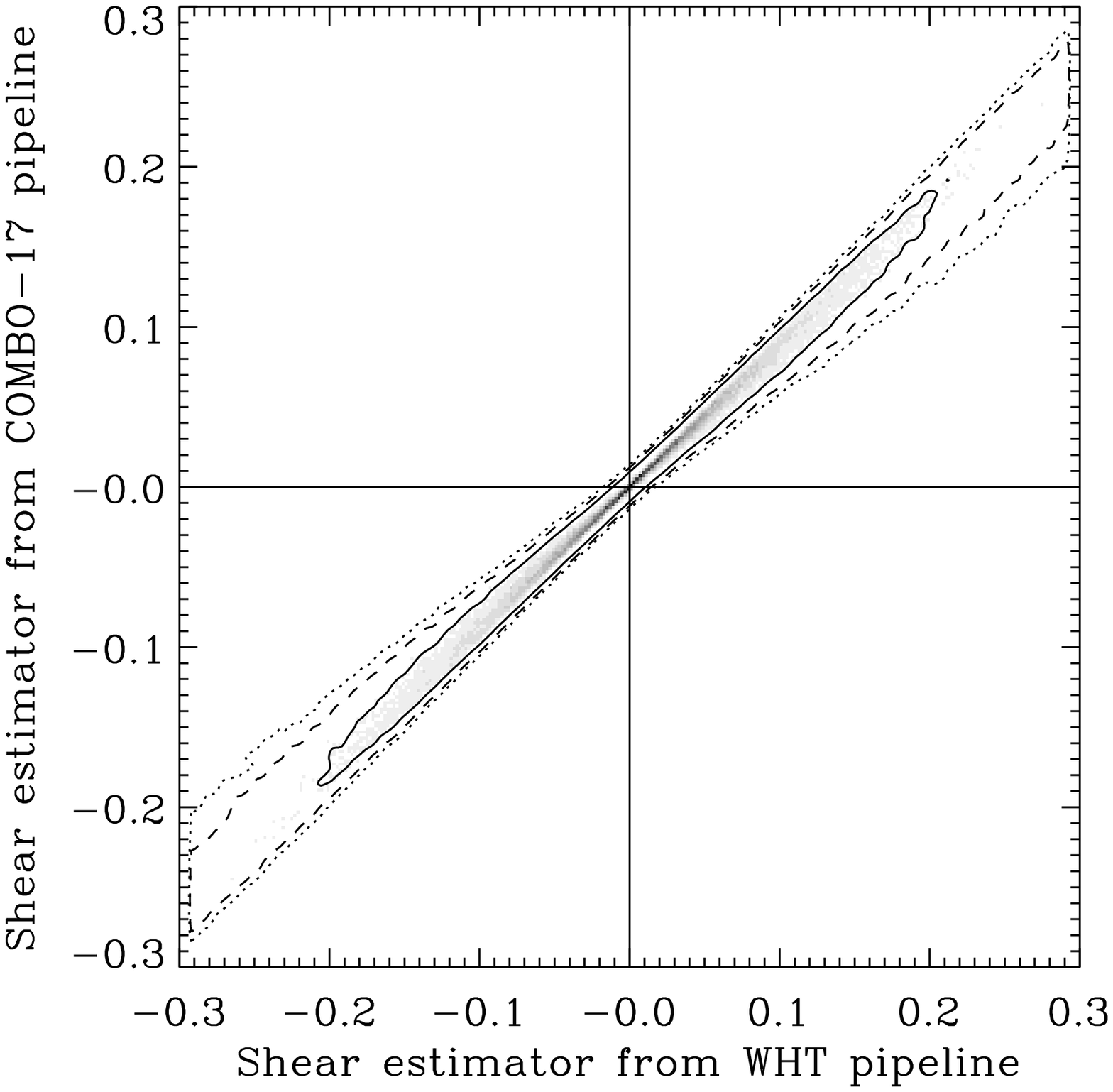,width=58mm,height=58mm}
\epsfig{figure=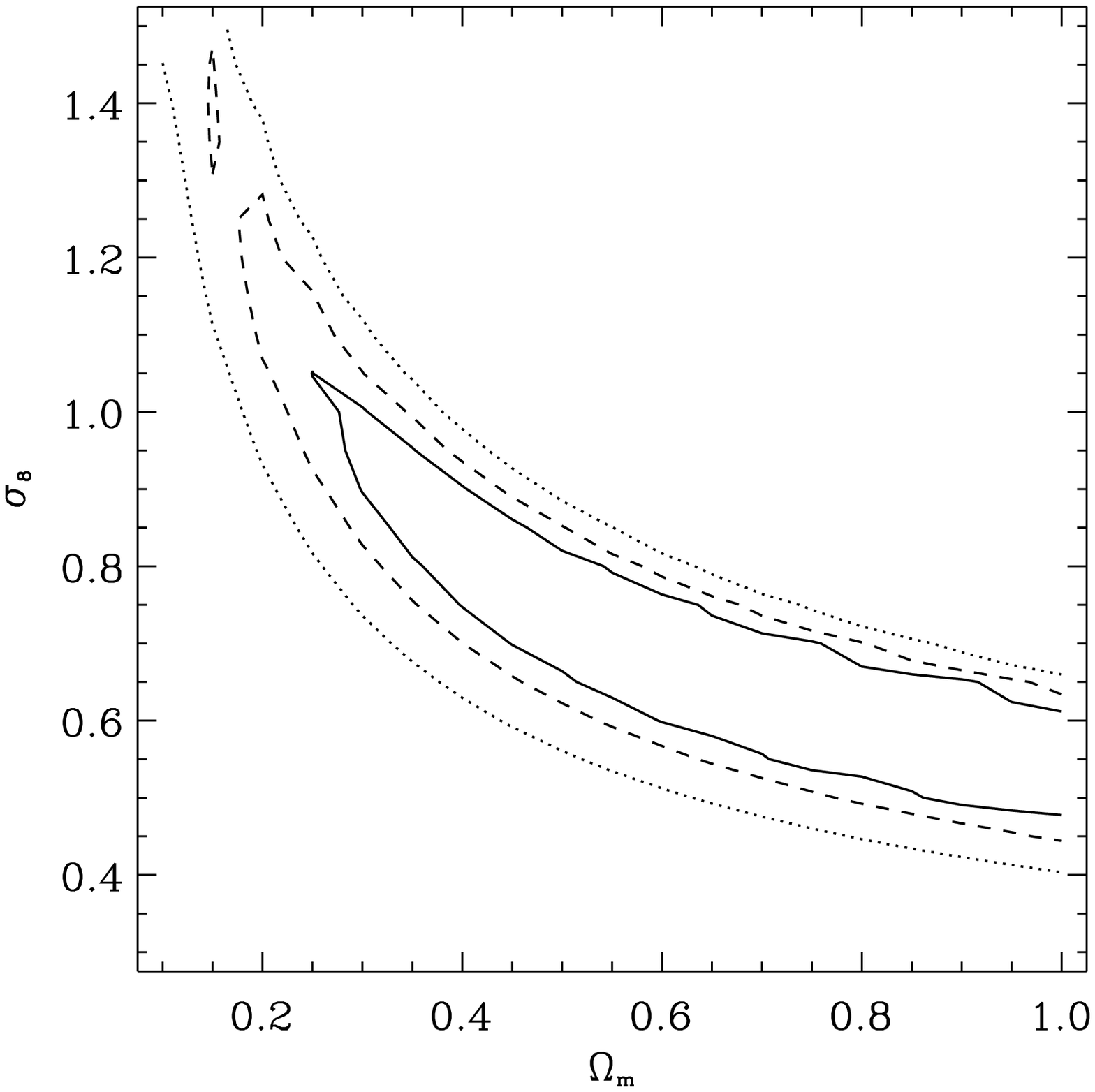,width=58mm,height=58mm}

\caption{{\it Left panels:} The shear susceptibility factor $P^\gamma$ for all
galaxies, as a function of their size $r_g$, with different options for the
signal-to-noise cut in the catalogue. The greyscale shows the number density of
galaxies throughout the parameter space, calculated by the COMBO-17 pipeline.
The WHT pipeline produces results almost identical to the top-left panel. In
practice the fit is performed on galaxies for one image at a time, and is
therefore more noisy; the fit to all galaxies here is shown merely to guide the
eye. The fit in the top panel is reproduced as a dashed line in the bottom
panel. When the COMBO-17 pipeline uses a signal to noise cut of $\nu>5$, the
population ensemble's average shear susceptibility is raised. {\it Middle
panels:} A comparison of the shear estimators derived via the WHT and the
COMBO-17 shear measurement pipelines. Only subsequently matched galaxies are
included in this plot, and the WHT pipeline always uses a signal to noise cut
to the catalogue of $\nu>15$, so all galaxies shown are brighter than that.
Consistent results are obtained in the top panel, where the COMBO-17
pipeline also uses a cut of $\nu>15$. The unbiased, 1\% dispersion of
individual values is merely due to the different interpolation of the PSF
across individual fields. When the COMBO-17 pipeline uses a signal to noise cut
of $\nu>5$, the shear estimators of all galaxies are lowered, including these
bright ones. {\it Right panels:} Constraints upon cosmological parameters 
$\Omega_m$ and $\sigma_8$, from a maximum-likelihood analysis of shears 
measured with the COMBO-17 pipeline, using different options for the
signal-to-noise cut. The 68.3\%, 95.4\% and 99.7\% confidence contours are 
shown.} 
\label{fig:analysis2} 
\end{figure*}

The mean components of shear across the entire CCD are $\langle \gamma_1
\rangle = (1.1\pm 7.1)\times 10^{-4}$ and $\langle \gamma_2 \rangle = (15.6\pm
7.0)\times 10^{-4}$. The rms shear within the survey is $\sigma_{\gamma_1} =
0.293$ and $\sigma_{\gamma_2} = 0.292$, or $\sigma_{|\gamma|} = \langle |
\gamma |^2 \rangle^{1/2} = 0.413$. (Note that our shear measurement pipeline
includes a catalogue cut at $|\varepsilon|<0.5$). The main, and irreducible,
component of this dispersion comes from the intrinsic ellipticities of source
galaxies. From other work performed with high resolution and high S/N
space-based data \citep{jr_stis,ref02} and simulated images \citep{snap2}, we
estimate a fundamental lower limit for $\sigma_{|\gamma|}$ around $0.30$.

\section{An independent analysis} \label{mlb}

Our measurement of $\sigma_8$ is at the relatively high end of the distribution
of published cosmic shear results. Compared to equivalently deep surveys,
it is most different from that published by the COMBO-17 survey by \citet{browncs}, 
who obtain $\sigma_8(\Omega_m/0.3)^{0.49}=0.72^{+0.08}_{-0.09}$. We have therefore
re-analysed our WHT data using several variations of the COMBO-17 pipeline
to determine the extent to which this disagreement might arise from technical
differences in the pipelines. 

There are four main differences between the COMBO-17 and the WHT pipelines.
In the COMBO-17 pipeline: 

\begin{itemize}

\item the PSF was interpolated across every field via third-order polynomials
in $x$ and $y$, so the correction may be different in idividual cases. Data was
excluded on all scales for any fields with problematic PSF correction,

\item a less stringent signal-to-noise cut was applied. All galaxies with
$\nu>5$ were included, rasising the overall number density to 19.8. Every
galaxy was still given the same weight when their shears were combined,

\item galaxy pairs were not included if the galaxies lay on different CCDs.
This will lower the signal-to-noise on large scales but may remove some bias if
the chips are imperfectly aligned on the focal plane,

\item no correction was made for astrometric distortions in the telescope. This
will spuriously increase the signal on large scales, where galaxies are a long
way from the telescope axis,

\item independently-developed code was used to calculate theoretical models and
to fit cosmological parameters to the data.

\end{itemize}

\subsection{Shear measurement}

The first two changes are relevant in the creation of a shear catalogue.
Differences in the PSF interpolation introduce a $\sim$1\% rms dispersion
between shear values measured by the WHT or COMBO-17 pipelines (see the
top-middle panel in figure~\ref{fig:analysis2}). This small difference can be
explained by the uncertain nature of any interpolation between sparsely sampled
data points. The lower-order fit in the WHT method typically produces smoother,
and therefore perhaps more reasonable behaviour in regions of the images that
are devoid of stars. However, neither method appears to bias the shear
estimators or resulting constraints on cosmological parameters. 

Of greater importance are the $\nu>15$ or $\nu>5$ catalogue cuts in
signal-to-noise. Very faint galaxies were discarded in our original analysis of
WHT images, and the simulated images of \citet{bacsim}, because of residual
correlation found between the galaxy and stellar ellipticities after
correction. Such correlations were not observed in the COMBO-17 data, so the
cut was lowered.  A possible explanation is the absence of CCD fringing in the
COMBO-17 data. Uncorrected fringing in WHT images would  alter the shapes of
those galaxies at a flux level comparable to the fringing via  an additive
process rather than a convolution or multiplication. This would not have  been
corrected, so the galaxies would have simply been discarded.  Brighter galaxies
that appear in both the WHT-pipeline and COMBO-17 pipeline  have identical
estimates of $P^\gamma$, since both are calculated by {\sc hfindpeaks}. 
However, these values are noisy. Both pipelines regard galaxies as a
population  ensemble, for which it is possible to average over noise
by fitting a global shear susceptibility factor.

Theoretically, the susceptibility factor of a galaxy shear is expected to vary
as a function of its radial profile, size and ellipticity; but not as a
function of its flux. We therefore fit $P^\gamma=P^\gamma(r_g)$, as shown in
the left-hand panels of figure~\ref{fig:analysis2}. However, it seems that real
faint galaxies do indeed have higher shear susceptibilities than
similarly-sized bright galaxies. As shown in the middle panels of
figure~\ref{fig:analysis2}, this raises $P^\gamma(r_g)$, and thus lowers shear
esimators by a factor of $0.85^{+0.12}_{-0.5}$. The skewed distribution
reflects the overall size distribtuion of galaxies, and arises because this
process affects small galaxies more than large ones. The reason for this
variation of $P^\gamma(\nu)$ is not yet clear: it may be that the morphology
distribution of faint galaxies really does contain intrinsically higher shear
susceptibilities. Both approaches would be valid in this case, since the fit
really has found a suitable average value for the population ensemble of
galaxies. 

A more worrying alternative would be that the noise in faint galaxies biases
$P^\gamma$ to higher values, partly because the main $P^{\rm sh}$ component of
$P^\gamma$ \citep[see][]{ksb} is defined to be strictly positive. In this case,
it would instead be preferable to fit only the bright galaxies, then apply that
susceptibility to faint galaxies of the same size. Either way, the results from
the WHT pipeline would be less affected, as the $\nu>15$ cut has been
calibrated upon simulated images containing a known signal by \citet{bacsim}.

A full investigation into this technical issue is beyond the scope of this
paper, and it will need further investigation in the future. Other comparative
studies, e.g. the Shear TEsting Program (STEP, Heymans \etal\ in preparation), 
or the Edinburgh and Bonn pipelines (Heymans \etal\ in preparation), find that 
the detailed way in which $P^\gamma$ is, or is not, adequately fit may indeed 
explain a large part of the variation between results from different
implementations  of KSB. Here, we shall propagate the analysis of shear
catalogues  from the COMBO-17 pipeline with both $\nu>15$ and $\nu>5$ cuts.

\subsection{Cosmological parameter constraints}

We find similar instabilities in the production of shear-shear correlation
functions with the COMBO-17 pipeline as we found in the WHT pipeline; results
are sensitive to binning at the $\sim\pm$7\% level. Measurements with the
$\nu>5$ as shown as blue circles in figure~\ref{fig:cth}, rescaled by a factor
$(0.8/0.9)^{1.65}$ to allow comparison despite the increased source redshift
distribtion. Excluding galaxy pairs of different CCDs has increased the size of
error bars by up to 50\% at large $\theta$, where the number of available
galaxy pairs is significantly lower, but the central values move only within
their $1\sigma$ error bars. Not correcting for the telescope's known
astrometric distortion has moved the bin at largest scales by a significant
amount, but this is not used for parameter fitting anyway. Excluding data on
all scales from fields with any residual star-galaxy correlation also has a
minimal effect. Each of these differences change the derived constraint on
$\sigma_8$ by less than 1\%.

The main difference in the two analyses is that introduced by the different
cuts in signal-to-noise. The inclusion of fainter galaxies increases the median 
magnitude of the $\nu>5$ population by $0.84\pm0.4$ magnitudes, leading to
a new median redshift of $z_s=0.90\pm0.07$ (\citet{cohen}). The expected cosmic 
shear signal thus increases (see equation~\ref{eq:coszgp}) --  an effect that is
incorporated during the calculation of theoretical models for the fitting of
cosmological parameters. Reassuringly, the shear signal measured from faint
galaxies is also higher. The final constraints for the COMBO-17 pipeline, using
a cut of $\nu>15$ indicate: 

\begin{equation} \sigma_{8} \left( \frac{\Omega_{m}}{0.3}
\right)^{0.52} = 1.01 \pm 0.13 ~, \label{eq:mlb_con_nu15} \end{equation}
including only statistical errors and, for a cut of $\nu>5$ with the higher
assumed source redshift distribution
\begin{equation} \sigma_{8}
\left( \frac{\Omega_{m}}{0.3} \right)^{0.52} = 0.98 \pm 0.10 ~.
\label{eq:mlb_con_nu5} \end{equation}
 
The consistency of these fits, and their agreement with the results from the 
WHT pipeline is reassuring. They tend to justify the adoption of a global 
shear susceptibility factor for an ensemble population of galaxies, and
verify the validity of the methods for controlling many potential systematic 
effects in both pipelines.

Our survey probes a wide range of angular scales, and thus excludes small
values of $\Omega_m<0.25$ (at 68\%CL). The preference for large $\Omega_m$
comes from the fit to the transfer function by \citep{bbks}. If we adopt the
fit by \citet{eisenhu}, and assume the Hubble parameter $h=0.7$, we effectively
apply a different prior on $\Omega_m$. In this case, our data excludes
$\Omega_m>0.55$ (at 68\%CL), similar to the result obtained by \citet{vw04}.
Using the \citet{eisenhu} transfer function also lowers our best-fit value for
$\sigma_8$ around $\Omega_m=0.3$ by $\sim$2\%.

\section{Conclusions} \label{conclusions}

We have measured the weak lensing shear-shear correlation functions in four
square degrees of deep $R$-band imaging data from the William Herschel
Telescope. Our measurements constrain the amplitude of the mass power spectrum,
$\sigma_8(\Omega_{m}/0.3)^{0.52}=1.02\pm 0.15$, including all contributions to
the total 68\%CL error budget: statistical noise, sample variance,  an
additional estimated error due to binning instabilities from non-Gaussian
outliers, covariance between different angular scales,  systematic measurement
and detection biases, source redshift uncertainty, and marginalisation with
priors over other parameters. We have examined our data for contamination by
systematic effects using a variety of tests including an $E$-$B$ decomposition.
These demonstrate a well-understood and modest contribution to our
uncertainties from systematic errors.

Using the pipeline developed for earlier WHT studies, we find our measurement 
of the normalization of the dark matter power spectrum lies at the relatively
high end  of the distribution of published values. However, it is still
consistent with those from  equivalently deep surveys by \citet{ref02} and
\citep{jr_stis}. Our results are also consistent at the 1$\sigma$ level  with
CMB results from the Wilkinson Microwave Anisotropy Probe (WMAP)
\citep{wmap1yr}.

The wide distribution of $\sigma_8$ constraints from recent cosmic shear
surveys understandably casts some aspersion upon their precision. For example,
it might be argued that the dispersion largely arises from unknown or
poorly-understood systematic effects. For the first time we analyze our dataset
with an independent pipeline -- that developed for the COMBO-17 data
(\citet{browncs}). This provides a valuable check on the extent to which
dispersion in the published cosmological results, as well as our apparently
high normalization, might arise from different techniques, both in constructing
shear catalogues and in analysing the shear-shear correlation functions.
Reassuringly, we find remarkable concordance between the two pipelines for the
same dataset. However, the various differences we have explored, for example in
the selection thresholds, seem to be insufficient to reconcile the spread in
observed $\sigma_8$ values, suggesting that significant differences remain at
some level in the actual data themselves.

Uncertainties in the redshift distribution of source galaxies in deep data
clearly contribute. It is difficult to determine the precise redshift
distribution of galaxies  after excluding those smaller than a fixed apparent
size. We have been conservative in this analysis and, as seen in
equation~(\ref{eq:con_full_errors}), source redshift uncertainty is already a
major component of our total error budget. The resolution of such issues will
require extensive spectroscopic follow-up and more complete image simulations.
Such advances are essential if the potential of the next generation of cosmic
shear surveys is to be fully realised.

Finally, we note that most recent cosmic shear results remain discrepant at the
$3\sigma$ level with measurements derived from the abundance of X-ray selected
cluster samples based on an observational rather than theoretical
mass-temperature relation \citep{bor,selclus,rei,vlnew}. These suggest
$\sigma_8\approx 0.75$. \citet{adamgauss} concluded that even extreme
non-Gaussianity in the mass distribution would be insufficient to explain this
discrepancy, because the two techniques probe similar mass scales. Further
studies are therefore needed in both the cluster method, to understand the
difference between the observed mass-temperature relation and that found in
numerical simulations; and in the weak lensing method, to construct more
reliable and better calibrated shear measurement methods. Such consistency
checks will represent a crucial verification of the standard $\Lambda$CDM
paradigm, so resolving this issue is of paramount importance.

\section*{Acknowledgements} We thank Neil O'Mahony, Chris Benn and the WHT
staff for their help with the observations. We thank Nick Kaiser for providing
us with the {\sc imcat} software, and to Douglas Clowe for advice on its use.
We thank Sarah Bridle, Catherine Heymans and Jason Rhodes for useful
discussions. We thank the anonynous referee for helpful suggestions to improve
the text and clarify various discussions. DB and MB were supported by PPARC
postdoctoral fellowships.

\bsp

\label{lastpage}

\end{document}